# Impact of a Deployed LLM Survey Creation Tool through the IS Success Model

Peng Jiang and Vinicius Cezar Monteiro de Lira and Antonio Maiorino

## Abstract

Surveys are a cornerstone of Information Systems (IS) research, yet creating high-quality surveys remains labor-intensive, requiring both domain expertise and methodological rigor. With the evolution of large language models (LLMs), new opportunities emerge to automate survey generation. This paper presents the real-world deployment of an LLM-powered system designed to accelerate data collection while maintaining survey quality. Deploying such systems in production introduces real-world complexity, including diverse user needs and quality control. We evaluate the system using the DeLone and McLean IS Success Model to understand how generative AI can reshape a core IS method. This study makes three key contributions. To our knowledge, this is the first application of the IS Success Model to a generative AI system for survey creation. In addition, we propose a hybrid evaluation framework combining automated and human assessments. Finally, we implement safeguards that mitigate post-deployment risks and support responsible integration into IS workflows.

## Introduction

Online survey platforms are integral to data-driven decision-making, offering organizations scalable mechanisms to gather insights on customer preferences, employee engagement, public health, market sentiment, and more. Viewed through an information systems (IS) lens, the survey platforms function as complex socio-technical ecosystems that collect, process, and transform human voice into actionable insights. Survey design and analysis is a great example of IS research studying how people, organizations, and technology interact to create value.

Despite their widespread use, designing effective surveys remains a significant challenge. Crafting questions that yield reliable and actionable data demands expertise in survey methodology, psychometrics, and domain-specific knowledge. One common issue is biased phrasing, such as double-barreled questions (e.g., "How do you like the food and service?"), which ask multiple issues and confuse respondents. Other hurdles include aligning questions with research goals, selecting appropriate question types—such as closed-ended questions (e.g., 'yes' or 'no') for quantitative data and open-ended questions for qualitative insights—and optimizing tone to engage target audiences.

In response, the rapid advancement of artificial intelligence (AI), particularly large language models (LLMs), offers a promising pathway to reimagine survey creation by integrating automation with human input. These technologies introduce new opportunities for digital integration, enabling scalable and theoretically grounded survey development while reducing cognitive and operational burdens. However, their deployment also raises critical questions about transparency, accountability, and security in the survey creation process.

This paper investigates the deployment and impact of an LLM-powered system for survey creation deployed in a widely used survey platform in late 2023. Users engage with the system by entering user prompts describing what they want to learn. The LLM interprets these prompts and generates relevant survey questions. To evaluate our LLM survey system's value, we apply the DeLone & McLean IS Success Model (DeLone and McLean, 2003) across four critical dimensions. First, we assess service quality by examining the support resources available to users, including documentation, help articles, and customer service channels. Second, we measure user satisfaction through built-in feedback tools that collect both numerical ratings and written comments. Using the Technology Acceptance Model (TAM) (Davis, 1989), we analyze this data to assess Perceived Ease of Use (PEOU), Perceived Usefulness (PU), and Behavioral Intention (BI), while also identifying key themes that shape user experiences. Third, we track the LLM system usage through comparative adoption metrics, contrasting our LLM tool's performance against four traditional survey creation methods available in our platform. We further analyze user prompts in the LLM system to understand real-world usage patterns. Finally, we propose a novel hybrid evaluation framework that combines expert human review with automated analysis to measure iterative system improvements. Survey researchers validate outputs for accuracy, relevancy, and ethical soundness, while the automated system tracks survey metadata and detects data drift to objectively compare quality across LLM versions. This is crucial for keeping pace with generative AI's rapid evolution and enabling data-driven decisions when adopting newer iterations of the LLM system. In parallel, we introduce deployment safeguards specifically designed to mitigate generative AI's unique risks, including system misuse, prompt leakage, and security vulnerabilities.

## Related Work

Information systems research has long explored data collection through technology. With advancements in AI, Collins (2021) offers a systematic literature review of AI's application in information systems research. The survey methods discussed in this paper are essential tools for data collection and analysis in information systems. Recent developments in LLMs have opened up new possibilities in this domain. For instance, Liang et al. (2025) introduced *SurveyX*, which leverages LLMs to enhance the quality and efficiency of academic survey development. Similarly, Wen et al. (2025) proposed *InteractiveSurvey*, which dynamically adapts question generation based on user interaction to personalize survey design. While these systems demonstrate impressive technical capabilities, they often lack grounding in established IS theoretical frameworks and overlook key considerations related to real-world deployment in survey applications.

However, established IS models have been available for over 20 years. They provide valuable frameworks for evaluating information systems, even though they have yet to be applied to the domain of survey generation. TAM model (Davis, 1989) provides insights into user adoption through perceived usefulness and ease of use. The DeLone and McLean IS Success Model (DeLone and McLean, 2003) offers a comprehensive framework for understanding user usage and evaluating system quality, information quality, and net benefits. Recent research has begun to apply the DeLone and McLean IS Success Model to evaluate AI-driven systems across various domains. For instance, Kulkarni et al. (2023) assessed the net benefits of generative AI systems in wealth management services, validating the success model's applicability in this context. Yoon & Kim (2023) explored improvement directions for AI speakers by applying the DeLone and McLean model, highlighting its relevance in consumer-facing AI technologies. Marjanovic et al. (2024) applied the IS Success Model to evaluate ChatGPT, demonstrating its utility in assessing the performance of generative AI tools. These applications illustrate the IS success model's versatility for evaluating emerging technologies.

However, to our knowledge, no prior work has applied the IS Success Model to evaluate LLM-powered systems for survey creation. Moreover, evaluating LLM systems presents unique challenges compared to traditional AI models, as generative tasks often lack clear ground truth. To address this, researchers have increasingly relied on human evaluations. For instance, Gómez-Rodríguez and Williams (2023) employed human raters to assess several LLMs on creative writing tasks. While effective, such evaluations are time-consuming and costly, underscoring the need for automated approaches. In the context of survey generation, where output is primarily text, the evaluation typically draws from metrics developed for assessing text data. Commonly used metrics include readability, length, and diversity, such as Flesch-Kincaid readability scores (Kincaid et al., 1975), Distinct-3 (Li et al., 2016), Self-BLEU (Papineni et al., 2018), and BARTScore (Yuan et al., 2021).

Our study contributes to the literature in three ways: First, we extend the application of the IS Success Model to a novel generative use case of an LLM-powered survey creation system, capturing the full lifecycle of system deployment and use. Second, we develop a hybrid evaluation framework that combines automated metrics with human assessments to evaluate the quality of the created surveys. Finally, we extend the human-AI collaboration by implementing post-deployment safeguards that address generative AI's unique risks, such as prompt leakage and security vulnerabilities, thereby enhancing the system's reliability and trustworthiness.

# A Deployed LLM System for Survey Creation

As the global leader in survey software, our platform collects over 20 million responses daily across diverse sectors, including healthcare, public policy, and market research. The survey process in our platform follows a structured workflow, as illustrated in Figure 1. The process begins with the crucial step of question formulation, where users design survey questions that effectively capture their research objectives while avoiding common pitfalls like biased phrasing or unclear wording. Following question creation, users enter a review phase where they can preview the survey's flow and content, making iterative refinements to ensure question validity and respondent comprehension. Once satisfied with the instrument, users proceed to share the survey through various channels to collect high-quality data. The final stage involves analytical processing of the gathered responses, where users transform raw data into actionable insights that inform organizational decision-making.

While AI could theoretically augment every workflow stage, this paper specifically examines how LLMs can transform survey creation, making it more accessible to non-experts while maintaining rigorous quality standards. We focus on the initial survey creation phase because it simultaneously presents the greatest adoption barrier and carries the highest stakes for outcome quality, where poorly constructed questions can systematically compromise data quality and downstream decision-making.

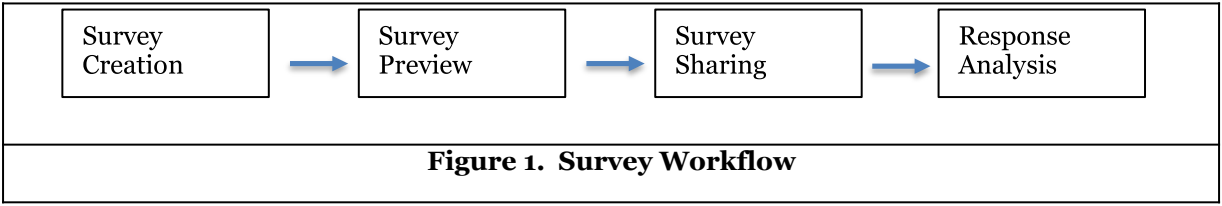

**Figure 1.  Survey Workflow**

The survey creation process begins with formulating an effective survey title that clearly communicates the instrument's purpose to potential respondents. More substantively, each survey question requires careful construction across three dimensions: the question text itself, the question type, and when applicable, the answer options.  The question text must be phrased precisely to avoid common pitfalls like leading language or ambiguous terminology. The **question type** determines whether responses will be structured (closed-ended) or unstructured (open-ended). **Closed-ended questions**, such as "What is your age range?" with predefined **answer options** (18-24, 25-34, etc.), enable efficient quantitative analysis but require exhaustive and mutually exclusive answer options. In contrast, **open-ended questions** like "Describe your experience in your own words" yield richer qualitative data but present greater analysis challenges.

Figure 2 illustrates the simplified architecture of our LLM System for survey creation, deployed in late 2023. The process begins when users provide descriptions about their survey objectives (e.g., "Create a customer satisfaction survey for an e-commerce platform"). We then use prompt engineering to enrich users' input, referred as user prompts, into a carefully crafted system prompt, incorporating survey design best practices and domain-specific constraints. The enriched system prompt guides the language model to generate high-quality surveys that align with users' stated goals.

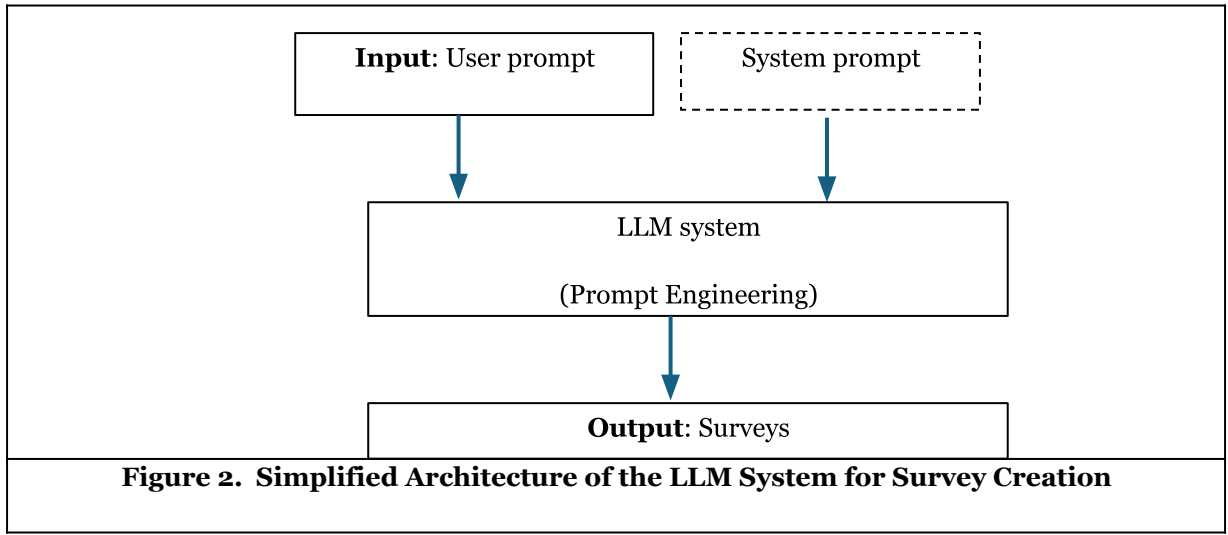

**Figure 2.** Simplified Architecture of the LLM System for Survey Creation

This architecture reflects our commitment to enhancing accessibility for non-expert users while preserving methodological rigor. The complexity of survey design is abstracted away through prompt engineering embedded in the system prompt provided to the LLM. Specifically, the system prompt performs several critical functions:

- **Prompt sanitization**: It filters out inappropriate or irrelevant user requests, such as those unrelated to survey design (e.g., mathematical problem solving).
- **Incorporation of survey design principles**: It encodes established best practices, including ordering questions from general to specific, deferring demographic questions to the end, and avoiding open-ended questions at the start to minimize early attrition.
- **Constraint enforcement**: It restricts question generation to predefined types supported by the platform, ensuring consistency and reliability.
- **Provide guidance on the selection of question types**: The system prompt encodes heuristics for determining when specific question types are appropriate. For instance, it advises that star rating questions, where respondents evaluate a statement using a five-point visual scale [1, 2, 3, 4, 5], should be employed judiciously, only when their visual affordances offer clear interpretive value.
- **Specify output formatting requirements**: It mandates that the generated survey be returned in a structured JSON format, aligning with production system requirements.

In the following section, we assess the effectiveness and value of the deployed LLM system using the DeLone and McLean IS Success Model, examining dimensions such as service quality, user satisfaction, system use, and information quality.

## Service Quality

To support user experience and ensure responsible use of the LLM system, the user interface provides a range of guidance and safeguards. These include character limits for input prompts, tips for writing effective prompts, reminders not to include sensitive information, and disclaimers about the potential inaccuracy of AI-generated content. Additionally, a help article is linked directly from the interface, offering clear information on the system's capabilities and limitations. This includes information on feature availability and configurations, supported question types, and constraints such as the maximum number of prompts allowed per hour.

To facilitate ease of adoption, we also provide a set of sample prompts covering a diverse range of use cases—including customer satisfaction, education experience, employee experience, event feedback, and market research—which serve as starting points for users to craft their own surveys.

For support and feedback, the system includes a feedback button that links to a dedicated survey where users can rate their overall experience and provide open-ended comments. In cases where technical issues arise, users are directed to contact the customer support team, who then triage and escalate issues to the relevant development teams for resolution.

## User Satisfaction

To assess user satisfaction with the LLM-powered survey creation system, we implemented a feedback survey accessible directly through the user interface. The survey includes two core questions: a closed-ended item measuring satisfaction on a 5-point Likert scale (1 = very dissatisfied, 5 = very satisfied), and an open-ended item inviting users to provide additional comments about their experience. A total of 557 responses were collected.

The average satisfaction score was 3.18 out of 5, indicating moderate user satisfaction. To gain deeper insights, we conducted qualitative analysis on the open-ended responses by segmenting users into two groups: promoters (ratings of 4 or 5) and detractors (ratings of 1, 2, or 3). This analysis demonstrates how open-ended user feedback can be systematically interpreted through the lens of the TAM. To extract structured insights from qualitative responses, we leveraged a second LLM to code each response for TAM constructs—Perceived Ease of Use (PEOU), Perceived Usefulness (PU), and Behavioral Intention (BI). We applied this process separately for the promoter group (users rating the system 4 or 5) and the detractor group (ratings of 1–3), enabling a comparative analysis.

Among promoters, 62% of responses referenced ease of use, with users highlighting that the system "removes the stress" of writing surveys and makes the process "fast and effortless." Another 55% cited perceived usefulness, describing the system as "very helpful," "saving hours of work," or producing "professional-quality surveys." Additionally, 48% of promoters expressed strong behavioral intention, noting they would "definitely use it again" or "highly recommend it to others."

Among detractors, common themes pointed to usability barriers rather than rejection of the concept. About 41% reported issues related to system unreliability, including long wait times. 34% mentioned UI design issues, such as fragmented navigation, where users must switch pages to see generated questions rather than allowing a seamless scroll. Editing friction was noted by 29%, including the lack of undo functionality after deleting questions. A further 25% raised transparency concerns, stating that they were unaware of the system-imposed limit on the maximum number of prompts that can be submitted within one hour. Several suggested that adding visual countdowns would be helpful.

It is worth noting that since many respondents identified multiple concerns in their feedback, the percentages exceed 100% due to overlap among themes. Overall, these findings provide actionable guidance: while promoters validate the system's core value and usability, detractor feedback points to friction points that can be addressed without fundamentally altering the system. Applying TAM in this context not only offers a structured framework for understanding user experience but also supports data-driven prioritization of improvements in future iterations of the LLM-powered system.

## System Use

In this section, we compare the usage of the LLM-based survey creation system with several traditional methods available on our platform. We also examine how users interact with the sample prompts provided in the LLM system interface and analyze user-generated prompts to uncover real-world usage patterns.

**Usage Comparison of LLM System with Other Survey Creation Methods**

Other than leveraging the LLM system for survey creation, referred to as *Build with LLM* in Table 1, our platform supports a few other ways to create surveys. The descriptions are provided in Table 1. In summary, users could create survey questions one by one or specify their requirements in a prompt, or leverage existing resources such as the survey templates with predefined questions provided by our platform, their previous surveys, or their questions written in external docs. Among all five methods, *Build with LLM* is the most recently introduced and the only one powered by AI.

| Survey Creation Method | Description |
| --- | --- |
| Build from Scratch | Manually create survey questions one by one |
| Build with LLM | Write a prompt and let LLM generate the survey |
| Copy from another survey | Copy survey questions from an existing survey and then update them as needed |
| Build from templates | Update surveys based on questions provided in survey templates |
| Import questions | Import survey questions from external sources such as Google Docs |

**Table 1. Description of Each Survey Creation Method**

We collect user engagement data over a year, from April 2024 to April 2025. Figure 3 shows that *Build from Scratch* is the most popular method users choose to create surveys. It has consistently been the dominant approach, even before *Build with LLM* was introduced. This trend is understandable, as many users may view survey creation as a relatively simple task that does not require AI assistance. Meanwhile, *Build with LLM*, as the most recently released and the only AI-powered feature, has quickly surpassed the other three traditional methods and become the second most-used option.

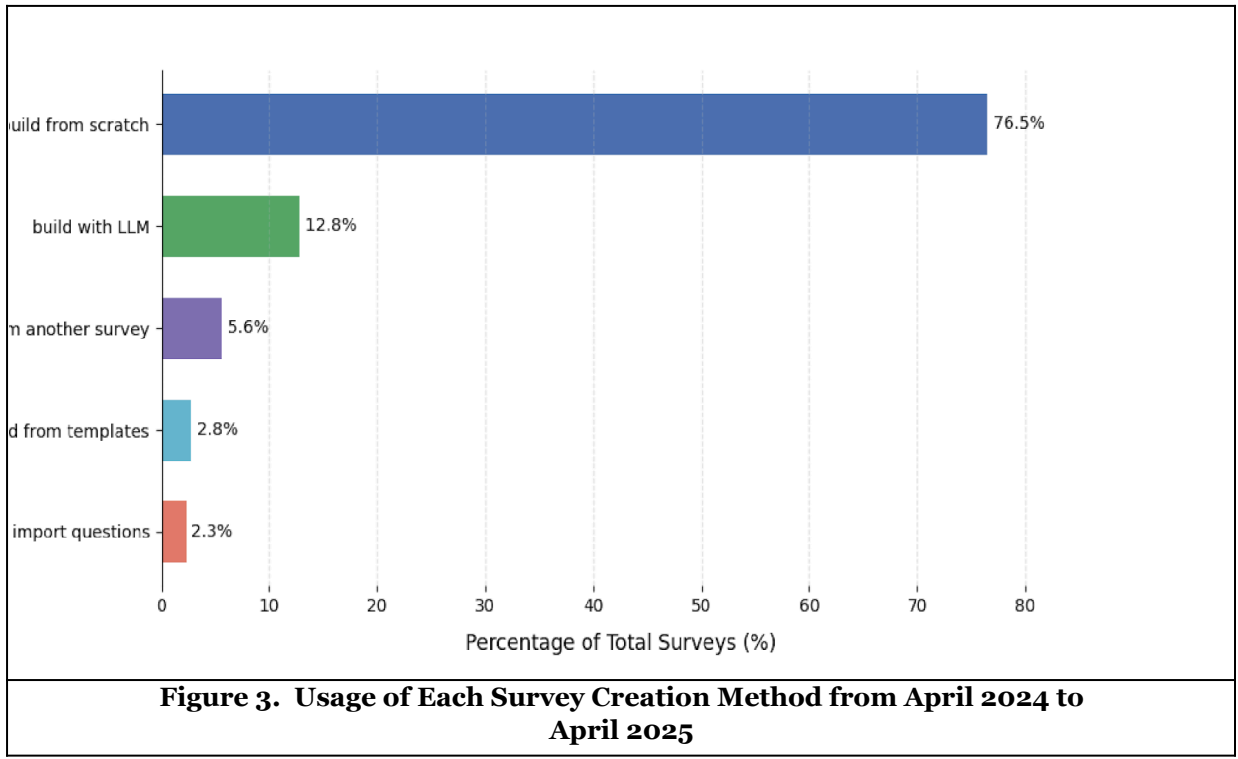

**Figure 3. Usage of Each Survey Creation Method from April 2024 to April 2025**

We further examine how each survey creation method supports user progress through the product workflow, as defined in Figure 1. Specifically, we evaluate two key indicators of survey success:

- **Activated**: the survey is shared with others
- **Deployed**: the survey receives at least five responses

Figures 3 and 4 present the activation and deployment rates for each survey creation method from April 2024 to April 2025. For example, surveys created via *Build with LLM* exhibit an activation rate of 48.6% and a deployment rate of 16.8%, indicating that out of every 100 surveys created using this method, approximately 49 are shared and 17 receive five or more responses.

Among all methods, *Copy from another survey* yields the highest activation and deployment rates. This is expected, as users leveraging this method are likely conducting repeated surveys in familiar contexts, such as longitudinal studies that track the same population over time. For instance, a study might monitor students' academic development across multiple years using consistent survey questions and respondent pools.

Notably, among the remaining four methods, *Build with LLM* achieves the highest activation and deployment rates. Although *Build from Scratch* remains the most widely used method overall, surveys created using *Build with LLM* are more likely to progress further in the workflow. *Build with LLM* demonstrates strong downstream value, helping more users move from initial survey creation to full survey deployment, ultimately contributing to greater net benefits in the real world.

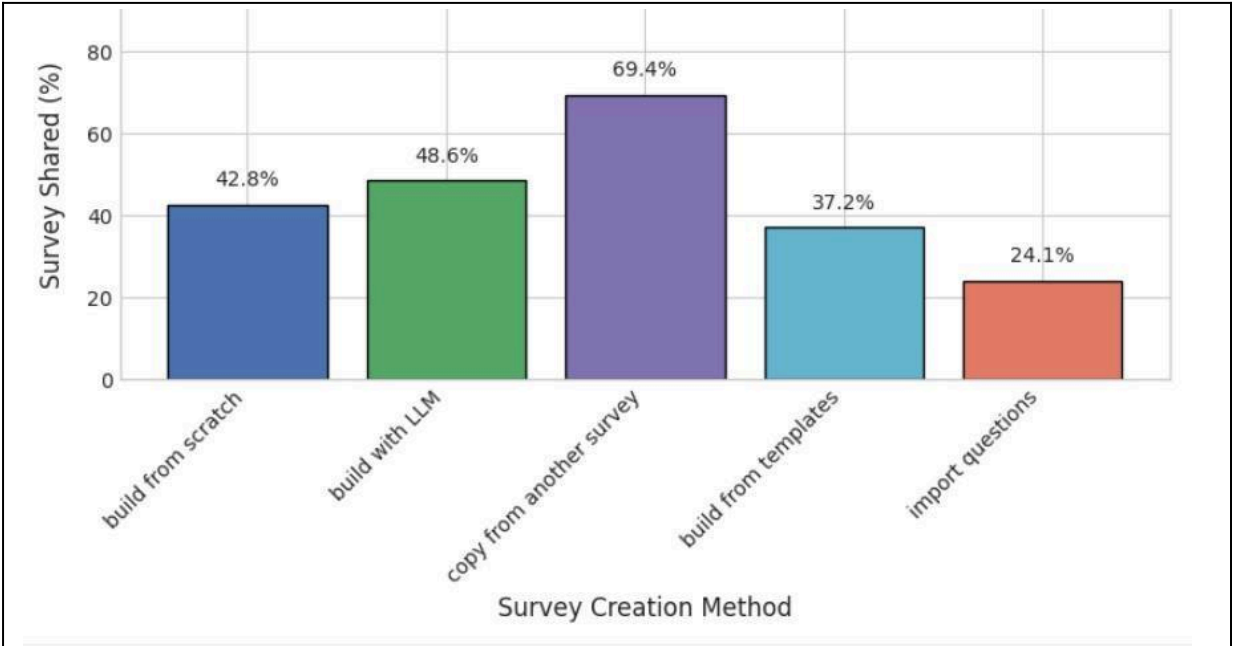

**Figure 4. Activation Rate of Each Survey Creation Method from April 2024 to April 2025**

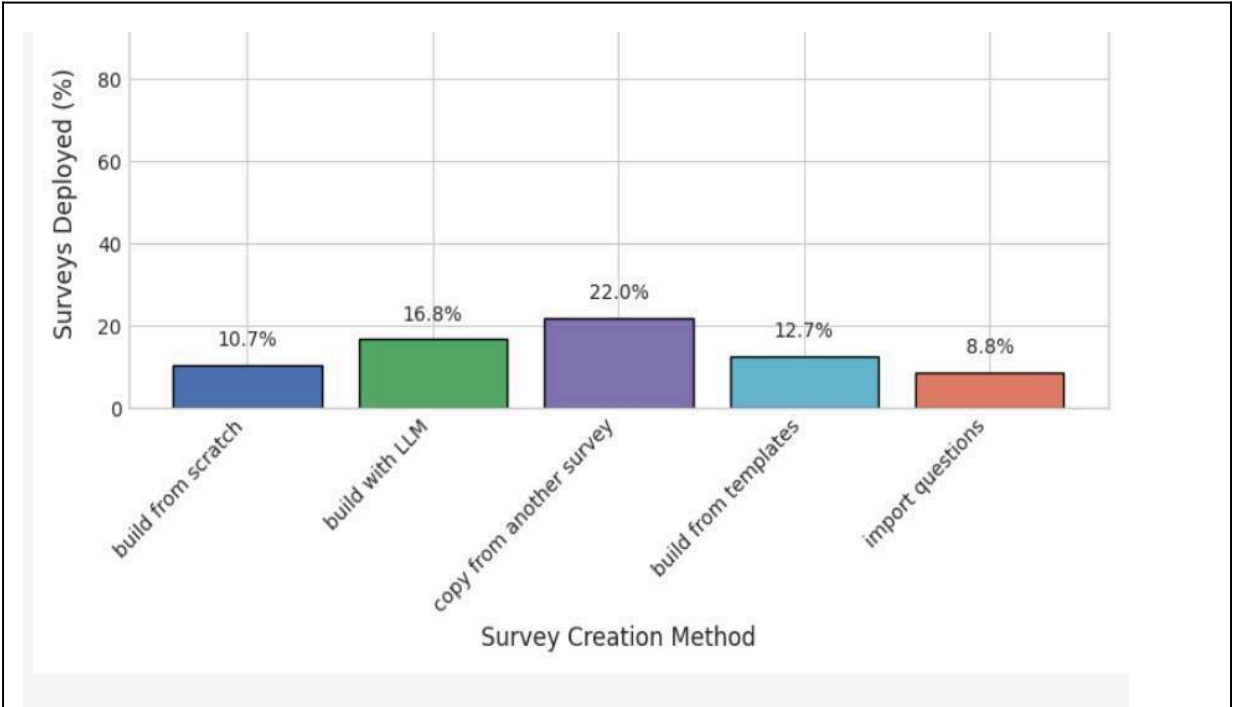

**Figure 5. Deployment Rate of Each Survey Creation Method from April 2024 to April 2025**

**Prompt Usage within the LLM System**

In addition to comparing the LLM system with other survey creation methods, we are also interested in examining how users interact with the suggested sample prompts and understanding the various use cases into which users' prompts fall. Table 2 presents sample prompts for five popular use cases, providing a jumpstart for different survey creation needs.

| Sample Prompts | Description |
| --- | --- |
| Customer satisfaction | A sample prompt for a survey assessing customer experience in areas like service quality, communication, etc. |
| Education | A sample prompt for a survey evaluating student satisfaction. |
| Employee experience | A sample prompt for a survey about the employee exit experience |
| Event feedback | A sample prompt for a survey collecting feedback on webinars |
| Market research | A sample prompt for a survey designed to gather insights about the market, like consumer behavior, purchase decisions, etc. |

**Table 2. Description of Sample Prompts**

Within the same one-year period, we found that approximately 37% of users utilized at least one of the sample prompts, while around 63% opted to write their own prompts. Figure 6 displays the usage of the sample prompts, with customer satisfaction, education, and market research being the three most frequently used categories among all five types.

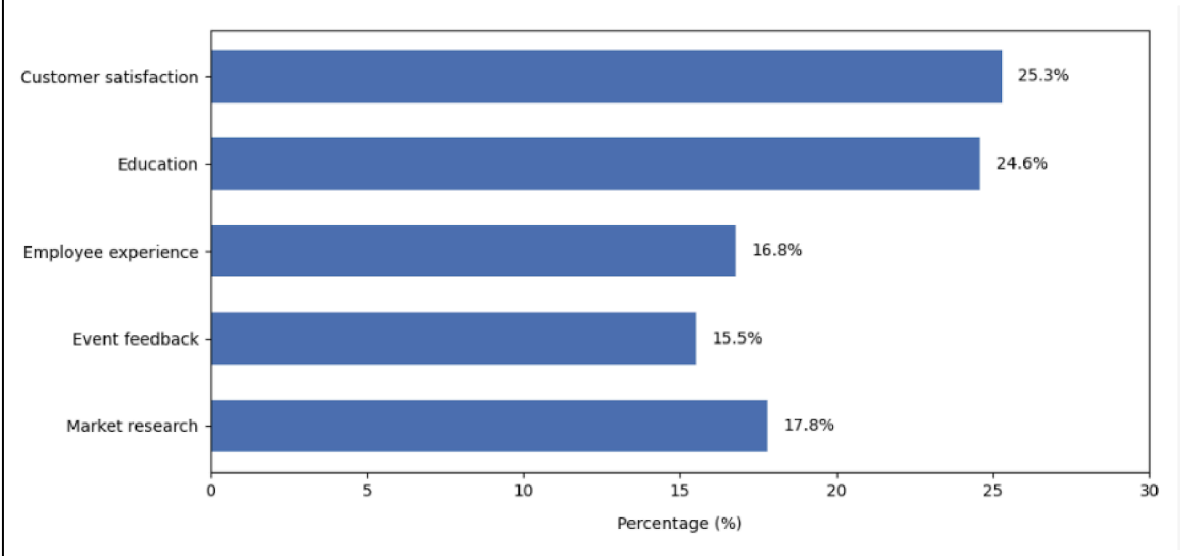

**Figure 6. Sample Prompt Usage from April 2024 to April 2025**

We further ran an in-house AI model to categorize user prompts into various use cases. Note that these user prompts include both the sample prompts and those created by users during the one-year period. Figure 7 presents the distribution of the various use cases of user prompts. A few insights can be drawn by comparing the categories in Figure 7 with those listed in our sample prompts in Table 2:

1. The categories in Figure 7 align with the five use cases outlined in Table 2, indicating that our sample prompts effectively cover a broad range of user needs.

2. Customer satisfaction remains the most frequently used category, consistent with its top usage among sample prompts, highlighting strong and sustained interest in this use case.
3. Two categories emerge: employee feedback and employee engagement. This suggests the opportunity to provide sample prompts tailored to more granular aspects of employee experience.
4. In addition to event feedback, a notable number of users express the need to generate surveys for event registration. This highlights a potential area for adding a sample prompt specific to this use case.
5. Lastly, the "Other categories combined" category in Figure 7 consists of a mix of lower-frequency use cases, such as votes, polls, and quizzes. While each contributes a small share individually, their presence signals diverse and evolving needs.

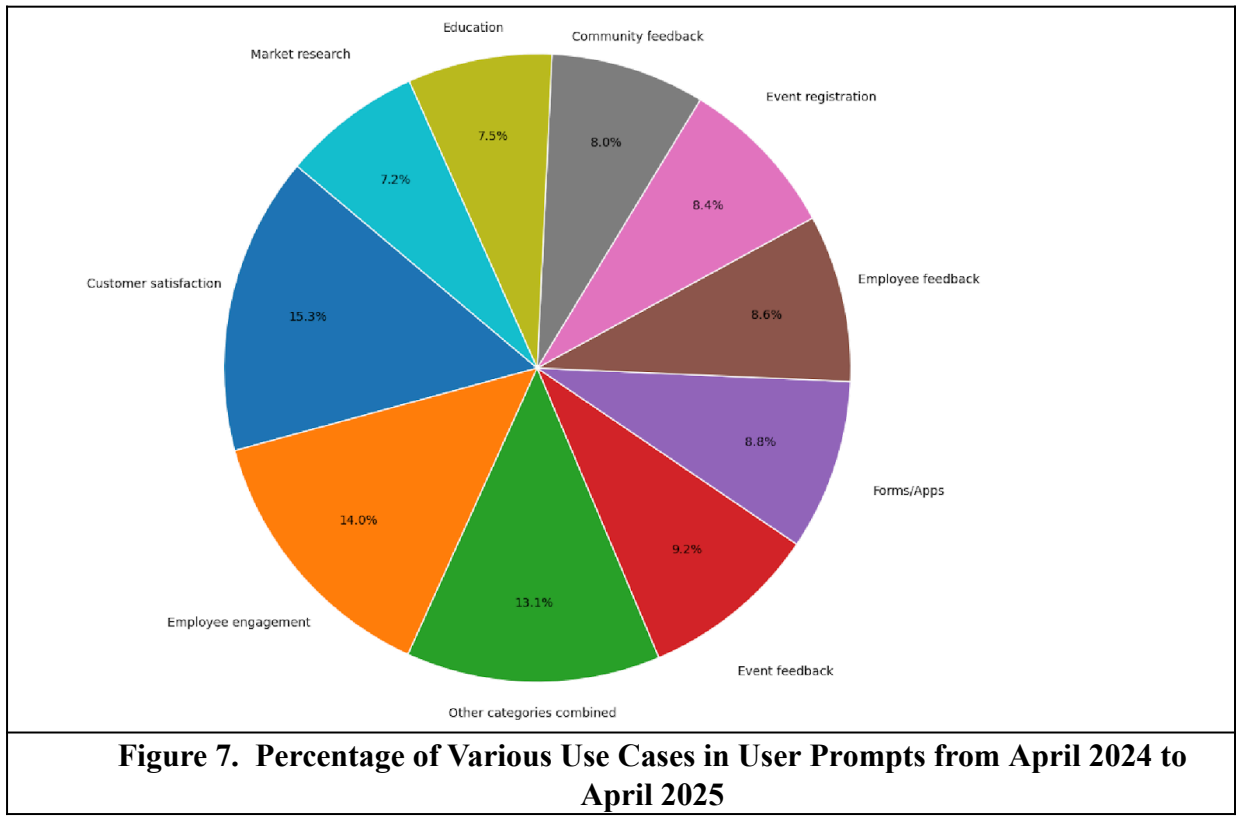

**Figure 7. Percentage of Various Use Cases in User Prompts from April 2024 to April 2025**

## Understanding User Behavior Through Binary Classification

Our LLM survey generation system begins with a user-provided prompt, based on which a draft survey is generated for preview. At this stage, the user can take one of three actions: (1) accept the survey and continue with downstream tasks such as sharing, (2) restart with a new prompt, or (3) exit the system. This subsection examines factors influencing a user's decision to accept a survey, as our goal is to advance users through the workflow.

We frame this as a binary classification task: for each prompt-survey pair, the outcome is either "accept" or "not accept." The latter category includes both restarting with a new prompt and dropping out. Table 3 summarizes the classification setup.

| Data Size | Class Balance Ratio (not accept: accept) | Model |
| --- | --- | --- |

| 234,300 prompt-survey pairs | 57:43 | LightGBM Ke et al. (2017) |

**Table 3. Information about the Binary Classification**

We apply stratified sampling to split the dataset into 80% for training and 20% for testing. The model uses features that capture metadata from the prompt, the generated questions, and user profile information such as industry and job role. Feature importance analysis reveals that prompt length, question length, and the number of answer options in the survey are the most predictive. Interestingly, user profile features are not significant. As shown in Figure 8, shorter prompts and more concise questions are more likely to lead to survey acceptance.

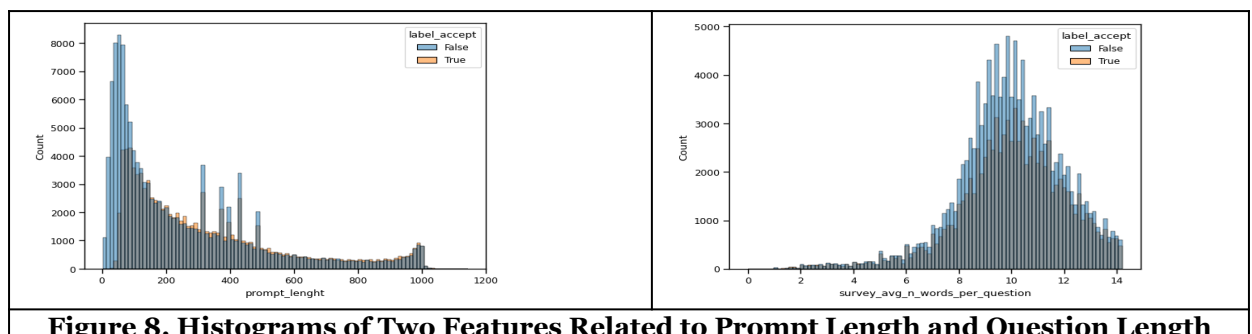

**Figure 8. Histograms of Two Features Related to Prompt Length and Question Length**

## Information Quality

To ensure the quality and consistency of generated surveys, we developed an evaluation framework for systematically assessing output across different versions of prompts and LLMs. This framework is critical for identifying the most effective configuration for production and for safeguarding against regressions as we evolve the system. The ultimate goal is to maintain a reliable user experience and prevent the release of system updates that might introduce unintended behavior.

Our LLM-powered survey creation system consists of two key components: (a) the system prompt, and (b) the LLM model. When the feature was first released in late 2023, we used GPT-3.5-Turbo (referred to as **GPT3.5**) in combination with the initial version of the system prompt (**Prompt V1**). Later, we explored GPT-4-Turbo (**GPT4**) alongside an enhanced prompt version (**Prompt V2**). The V2 prompt introduced several improvements, such as support for multilingual survey generation (beyond English), clearer output formatting instructions, and guidance to generate a broader variety of question types, reducing the dominance of open-ended questions seen in the system prompt V1. Both GPT3.5 and GPT4 used in this study were accessed via the "0125" release of the OpenAI API. Altogether, we evaluated four combinations, two LLM variants combined with two prompt versions, to determine the best-performing setup.

**Data Collection for LLM System Evaluation**

The user prompts used for this evaluation were collected from customers who interacted with the LLM system between October 2023 and January 2024, as the evaluation was conducted at that time. To ensure data quality and relevance, we applied the following filtering steps in sequence:

1. Remove duplicate prompts.
2. Exclude prompts flagged by our internal privacy-preservation pipelines as containing PII or sensitive information.
3. Select only prompts written in English.

4. We excluded prompts that were either unusually short or long (i.e., <200 or >500 characters), or whose generated surveys contained an atypical number of questions (i.e., <5 or >12), based on distribution analyses. These filters ensured that our evaluation focused on representative user behavior and typical survey outputs.

After applying these filters, we obtained a high-quality set of user prompts for evaluating different combinations of LLMs and system prompts. However, defining a correct survey for each prompt is inherently difficult. Unlike traditional machine learning tasks that rely on clearly defined ground truth, generative tasks like survey creation lack such definitive answers. To address this challenge, our evaluation framework combines human judgment with automated metrics to assess survey quality more holistically.

**Human Evaluation of the LLM System**

For the human evaluation, we collaborated with survey research experts to define a set of metrics that assess how well a generated survey aligns with the intent of the user prompt. These metrics, summarized in Table 4, are organized into two categories: question-level and survey-level. Each metric is scored on a scale from 0 to 2, with higher scores reflecting better quality.

At the question level, the evaluation focuses on aspects such as clarity of wording in the question texts, the relevance of answer options, and the presence of biased questions. At the survey level, metrics assess whether the survey sufficiently addresses the prompt's requirements, maintains overall relevance to users' prompts, and avoids being overly dominated by a single type of question.

| Metric | Question-level | | | Survey-level | | |
|---|---|---|---|---|---|---|
| | Question text quality | Answer options | Bias check | Missing questions | Relevance to prompt | Question variety |
| Scoring | 0 = poor quality/strange wording<br><br>1 = some issues but OK<br><br>2 = good/acceptable wording | 0 = irrelevant/too many<br><br>1 = mostly relevant/good number of answer options<br><br>2 = well-tailored/ideal number of answer options | 0 = if many biased questions are generated<br><br>1 = if limited<br><br>2 = all questions are neutral | 0 = if something explicitly asked in the prompt was not included in the survey<br><br>1 = if something from the prompt is missing, but is indirectly covered by other questions<br><br>2 = if the survey includes questions explicitly asked in the prompt | Check whether the questions in the survey are relevant to the prompt issued:<br><br>0 = completely irrelevant<br><br>1 = relevant but not tailored<br><br>2 = completely relevant | Check whether the survey has too many open ended questions:<br><br>0 = too many open-ended questions<br><br>1= somewhat repetitive<br><br>2 = varied types of questions |

**Table 4. Evaluation Checklist for Survey Evaluation**

When we make updates to the LLM system, our partners from the survey research team manually score the generated surveys. The human evaluation allows us to incorporate domain experts' knowledge to assess survey quality. However, this process is resource-intensive and time-consuming. To complement it, we also developed an automated evaluation framework. In a nutshell, the automated evaluation begins by defining a list of features that capture key attributes of the generated surveys, and then monitors changes in these features over time or conducts drift tests to detect shifts when updates are made to the LLM system. We will go through the details of the automated evaluation method in the next subsection.

**Automated Evaluation of the LLM System**

Table 5 outlines the survey metadata features used in our evaluation framework. The first column indicates the name of each feature, while the second specifies the aggregation function applied. For example, the feature *n_generated_questions* represents the total number of questions in a survey. Most feature names are self-explanatory. They capture counts of different question types (e.g., open-ended, NPS, multiple selection), as well as statistics related to the length of question text and answer options (e.g., average number of words per question, average word length, etc.).

In addition, the last five features are inspired by Robust Intelligence, a platform for testing and monitoring AI models, to help assess the overall quality of the generated text. Specifically, we concatenate all question texts in a survey into a single string and compute the following:

- *any_special_character*: A Boolean flag indicating whether special characters such as "^", "&", or other non-standard punctuation appear in the text.
- *score_flesch_kincaid* scores (Kincaid et al., 1975): A readability score based on the Flesch-Kincaid formula, which estimates how easy a text is to read.
- *dist_unigrams* and *dist_bigrams*: The distribution of unigrams and bigrams in the concatenated text, used to monitor lexical variety and detect potential drift in vocabulary patterns.
- dist_character: The distribution of characters in the concatenated text, like letters and numbers.

| Feature Name | Aggregation |
|---|---|
| n_generated_questions | count |
| n_open_ended_questions | count |
| n_closed_ended_questions | count |
| n_multiple_selection_questions | count |
| n_single_choice_questions | count |
| n_contact_info_questions | count |
| n_nps_questions | count |
| n_unsupported_questions | count |
| n_characters_in_survey | count |
| n_words_in_survey | count |
| std_n_words_per_question | std |
| avg_word_length_in_survey | mean |
| avg_n_answer_options | mean |
| avg_n_words_per_question | mean |
| avg_n_words_per_answer_option | mean |
| max_word_length_in_survey | max |
| any_special_character | count |
| score_flesch_kincaid | count |
| dist_unigrams | count |
| dist_bigrams | count |
| dist_character | count |

Table 5. List of Survey Metadata Features

We can then use the Population Stability Index (PSI) introduced by Taplin and Hunt (2019) to assess the distributional shifts in feature values between different implementations of the LLM system. The PSI is a widely used metric to track changes in distributions over time, helping us understand how updates to system prompts and LLM models affect the generated surveys. To compare the feature values of generated surveys corresponding to two different LLM systems, we compute the PSI for each feature value to detect any distributional shifts between the implementations. The typical interpretations of PSI outcomes are as follows:

- PSI < 0.1: Indicates no significant population change.
- PSI < 0.2: Reflects a moderate population change.
- PSI ≥ 0.2: Signifies a significant population change.

Any PSI value above 0.2 is called a FAILED test for the feature, indicating significant changes in the distribution of the feature. Table 6 lists the PSI values for a feature in the four comparisons with the two variants of system prompts and two LLM models. Note that we include only the features that failed in at least one of the four comparisons. The number marked in bold indicates the feature that has the most significant change during a comparison. For example, in the first column, when comparing system prompt V1 and V2, the number of multiple selection questions has the most noticeable change. This is because the system prompt V2 asks the LLM to create more closed-ended questions, whereas V1 tends to generate too many open-ended questions. Multiple selection questions are a popular type of closed-ended questions, allowing respondents to choose more than one option from a list of possible answers. Figure 9 shows clearly that the V2 prompt tends to generate more multiple selection questions than the V1 prompt, resulting in a broader variety of question types in the surveys.

| Features | $<\mathcal{B}_{v1}^{GPT3.5}, \mathcal{B}_{v2}^{GPT3.5}>$ | $<\mathcal{B}_{v1}^{GPT3.5}, \mathcal{B}_{v1}^{GPT4}>$ | $<\mathcal{B}_{v2}^{GPT3.5}, \mathcal{B}_{v2}^{GPT4}>$ | $<\mathcal{B}_{v1}^{GPT4}, \mathcal{B}_{v2}^{GPT4}>$ |
|---|---|---|---|---|
| avg_n_answer_options | 0.405 | 0.478 | 0.424 | 0.351 |
| avg_n_words_per_answer_option | 0.642 | 0.311 | 0.339 | 0.622 |
| avg_n_words_per_question | 0.000 | 0.526 | 0.392 | 0.367 |
| drift:bigrams_distribution | 0.565 | 0.860 | 0.795 | 0.673 |
| drift:unigrams_distribution | 0.205 | 0.247 | 0.222 | 0.217 |
| max word length | 0.000 | 0.000 | 0.219 | 0.000 |
| n_closed_ended_questions | 0.000 | 0.318 | 1.203 | 0.447 |
| n_contact_info_questions | 0.000 | **3.778** | 0.529 | **1.540** |
| n_generated_questions | 0.000 | 2.624 | **2.950** | 0.000 |
| n_multiple_selection_questions | **0.991** | 0.000 | 0.395 | 1.271 |
| n_nps_questions | 0.000 | 1.899 | 2.300 | 0.000 |
| n_open_ended_questions | 0.000 | 0.352 | 0.846 | 0.000 |
| n_single_choice_questions | 0.000 | 0.000 | 0.504 | 0.000 |
| n_words_in_survey | 0.000 | 1.535 | 2.068 | 0.000 |
| std_n_words_per_question | 0.000 | 1.173 | 0.419 | 0.599 |
| n_characters_in_survey | 0.000 | 1.446 | 1.871 | 0.000 |

**Table 6. PSI Score per Feature**

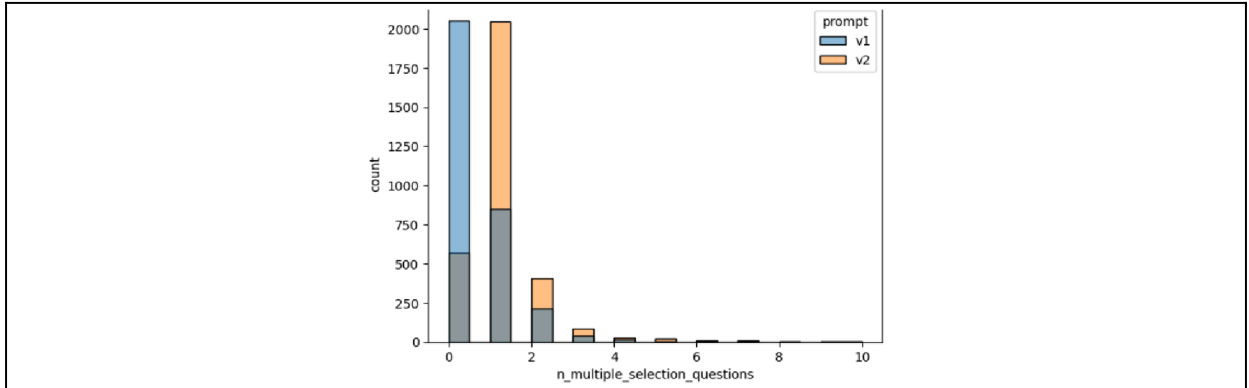
**Figure 9. Histograms of *n_multiple_selection_questions* using GPT3.5 with V1 and V2**

After calculating the Population Stability Index (PSI) scores for each feature across comparisons, we aggregate the results in Table 7. Recall that a failed PSI test indicates statistically significant divergence in feature distributions between model versions. Our analysis reveals a pronounced increase in failed tests when transitioning from GPT3.5 to GPT4: 13 failures under system prompt V1 and 16 under V2. This consistent pattern across prompt versions demonstrates GPT4's fundamentally different survey generation behavior compared to GPT3.5.

When triangulated with human evaluation results, these quantitative findings informed our production deployment decision: GPT4 with system prompt V2 was selected for its superior balance of question diversity and coherence, as evidenced by both evaluation metrics and expert assessments.

| Experiment | FAIL | PASS |
|---|---|---|
| $<\mathcal{B}_{v1}^{GPT3.5}, \mathcal{B}_{v2}^{GPT3.5}>$ | 5 | 16 |
| $<\mathcal{B}_{v1}^{GPT3.5}, \mathcal{B}_{v1}^{GPT4}>$ | 13 | 8 |
| $<\mathcal{B}_{v2}^{GPT3.5}, \mathcal{B}_{v2}^{GPT4}>$ | 16 | 5 |
| $<\mathcal{B}_{v1}^{GPT4}, \mathcal{B}_{v2}^{GPT4}>$ | 9 | 12 |

**Table 7. Overall Results of Drift Tests in Each Comparison**

**Post-Deployment Safeguards in the LLM System**

Following the selection of GPT4 with system prompt V2, we continue to monitor the deployed system to safeguard information quality and system integrity. We establish a rigorous evaluation process leveraging dedicated resources to identify potential vulnerabilities and edge cases that automated metrics may overlook. Their findings inform prompt updates that address three critical areas:

First, the system now filters out inappropriate user requests, such as solving math problems or completing coding tasks, ensuring that it remains focused solely on survey generation. Second, we strengthen protections against prompt leakage, a known security risk in which LLMs unintentionally reveal system instructions. Based on adversarial prompts identified through our internal evaluation process, we revise the system prompt to recognize and reject such attempts. Third, we implement rate limiting by setting a cap on the number of prompts a user can submit per hour. This measure reduces the risk of denial-of-service (DoS) attacks, in which malicious actors attempt to overwhelm the system and prevent legitimate users from accessing it, thereby helping maintain overall system reliability and availability.

These ongoing refinements reflect our commitment to quality across content relevance, security, and system stability. The human-in-the-loop evaluation process proves essential in surfacing issues beyond the scope of automated tools, reinforcing the importance of continuous human-AI collaboration in the safe deployment of generative systems.

## Conclusion

In this study, we presented a deployed LLM-powered system designed for survey creation. We evaluated its value through the lens of the DeLone and McLean IS Success Model, assessing key dimensions including service quality, user satisfaction, system use, and information quality. Since deployment, the system has achieved strong adoption, outperforming several traditional survey creation methods offered on our platform. The TAM model results further indicate that users find the tool easy to use, helpful, and worth recommending to others. This success is reflected in downstream metrics such as higher survey sharing rates and improved response collection. Moreover, we observed diverse usage patterns, with the top three applications centering on customer satisfaction, employee experience, and event feedback.

As LLM technology evolves rapidly, our comprehensive evaluation framework plays a critical role in maintaining a stable, safe, and reliable survey creation experience. By integrating both automated and human-in-the-loop evaluations, the framework allows us to iteratively refine system prompts and model configurations. It also serves as a protective barrier, setting essential guardrails to prevent unexpected behaviors and reinforce system integrity during updates.

In sum, this work showcases a seamless AI-powered digital integration within a real-world product. It highlights the potential of LLM-powered tools to shape the future of survey design, aligning with the IS tradition of examining how people and technology co-create value. Our findings point to several promising directions for future improvements: offering more sample prompts tailored to users' top use cases, improving system latency, and enhancing UI transparency to further elevate perceived service quality.

## References

Collins, C., Dennehy, D., Conboy, K., & Mikalef, P. (2021). *Artificial Intelligence in Information Systems Research: A Systematic Literature Review and Research Agenda. International Journal of Information Management, 60*, 102383. https://doi.org/10.1016/j.ijinfomgt.2021.102383

Davis, F. D. (1989). *Perceived Usefulness, Perceived Ease of Use, and User Acceptance of Information Technology. MIS Quarterly*, 13(3), 319–340.

DeLone, W. H., & McLean, E. R. (2003). *The DeLone and McLean Model of Information Systems Success: A Ten-Year Update. Journal of Management Information Systems*, 19(4), 9–30.

Gómez-Rodríguez, C., & Williams, P. (2023). *A Confederacy of Models: A Comprehensive Evaluation of LLMs on Creative Writing*. In *Findings of the Association for Computational Linguistics: EMNLP 2023* (pp. 14504–14528). Association for Computational Linguistics. https://doi.org/10.18653/v1/2023.findings-emnlp.966